\documentclass[10pt]{article}
\usepackage{amsfonts}
\usepackage{amsmath}
\usepackage{amssymb}
\usepackage{epsfig}
\usepackage{graphicx,color}
\usepackage{graphicx}
\def\sl{\!\!\!\slash}
\begin{document}
\title{Impact of Family Non-universal $Z^\prime$ Boson on  Pure Annihilation $B_s \to \pi^+ \pi^-$ and $B_d \to K^+ K^-$ Decays}
\author{Ying Li\footnote{Email:liying@ytu.edu.cn}, Wen-Long Wang, Dong-Shuo Du, and Zuo-Hong Li\\
{\it Department of Physics, Yantai University, Yantai 264-005, China}\\
Hong-Xia Xu \\
{\it Wenjing Colledge, Yantai University, Yantai 264-005, China}
}
\maketitle
\begin{abstract}
We study the $B_s \to \pi^+ \pi^-$ and $B_d \to K^+ K^-$ decays in the standard model  and the family non-universal $Z^\prime$ model. Since none of the quarks in final states is the same as the initial quark, these decay modes can occur only via   power-suppressed annihilation diagrams. Despite the consistence of the standard model prediction with the available data, there is  a   surviving  room    for a light $Z^\prime$ boson.  Taking into account  the $Z^\prime$  contribution, we find  theoretical results  for  branching fractions can better accommodate the data. With the relevant data, we  also derive a constraint on the parameter space for the $Z^\prime$.  Moreover, for the $B_d \to K^+ K^-$,  both the direct and the mixing-induced $CP$ asymmetry are sensitive to the couplings between $Z^\prime$ and fermions in the parameter spaces constrained by data. The measurements at future experimental facilities, including the LHC-b, Belle-II and the proposed high energy $e^+e^-$ collider, will provide us useful hints for direct searching for the light $Z^\prime$ boson.
\end{abstract}
\section{Introduction}\label{sec:1}
Since the  discovery of the Higgs boson at the Large Hadron Collider (LHC) \cite{higgs}, the search for new physics (NP) degrees of freedom beyond Standard Model (SM)  becomes  one of the most important tasks of high energy particle physics. In many NP models, an extra $U(1)^\prime$ gauge symmetry is often introduced based on various motivations in new physics beyond SM, resulting in an additional massive neutral gauge boson usually called the $Z^\prime$ boson.  Quite a few models are of this type,  such as grand unified theories based on the gauge groups SO(10) \cite {so10}, $E_6$ model \cite{e6}, supersymmetric models \cite{susy}, and string inspired models \cite{string} (for a review, see Ref.~\cite{Langacker:2008yv}).  Although the $U(1)^\prime$ charges are usually family-universal,  it is not mandatory to be so,  and the family non-universal $Z^\prime$ has been introduced in some models, such as in aforementioned $E_6$ model \cite{e6}.

On the experiment side, many efforts have been expended  to search for the $Z^\prime$ directly at the LEP, Tevatron, and LHC. With the assumption that the $Z^\prime$ couplings to the SM fermions are similar to  those of the SM $Z$ boson, the direct searches for the $Z^\prime$ can be performed in the dilepton events. At this stage, the lower mass limit has been set as $2.86~\mathrm{TeV}$ at the $95\%$ confidence level (CL) from collisions at 8 TeV with an  integrated luminosity of $19.5 \mathrm{fb}^{-1}$ by using $e^+e^-$ and $\mu^+\mu^-$ \cite{ATLAS:2013jma} events, and this value becomes  $1.90~\mathrm{TeV}$  using the   $\tau^+\tau^-$ events~\cite{TheATLAScollaboration:2013yha}.

However, if the $Z^\prime$ boson does not couple to leptons, the above constraint from the LHC is no longer valid. Theoretically, such leptophobic $Z^\prime$ boson can be realized in $E_6$ model \cite{e6}, the phenomenological studies at the LHC has been recently explored in Ref.~\cite{Chiang:2014yva}. In complementary to the direct search, some characters of the leptophobic $Z^\prime$ boson can also be constrained indirectly from the ``low" energy flavor physics. The family non-universal $Z^\prime$ boson may induce tree-level flavor changing neutral currents (FCNC) and thus they are severely bounded  by experiment, most notably meson mixing \cite{fcncconstraints}. Other effects of the FCNC in flavor physics have also been studied in past decades \cite{chiang1,BZprime,changkllzp}.  Motivated by the above arguments, we aim to in this work perform a comprehensive analysis of the impact of a family non-universal $Z^\prime$ boson on the pure annihilation decays $B_d \to K^+K^-$ and $B_s \to \pi^+ \pi^-$. Since these modes are power suppressed in the heavy quark limit, their branching ratios are expected to be very small, and the sensitivity to NP can be then enhanced.

Experimentally, the decay mode $B_s \to \pi^+ \pi^-$ was firstly reported by the CDF collaboration
\begin{eqnarray}
{\cal B}(B_s \to \pi^+ \pi^-)&=(0.57 \pm 0.15 \pm 0.10) \times 10^{-6}~\cite{Aaltonen:2011jv} ,
\end{eqnarray}
and it was soon confirmed by the LHCb collaboration with $0.37~\mbox{fb}^{-1}$ data as
\begin{eqnarray}
{\cal B }(B_s \to \pi^+ \pi^-)&=(0.95^{+0.21}_{-0.17} \pm 0.13) \times 10^{-6}~\cite{Aaij:2012as} .
\end{eqnarray}
So, the averaged result is given as~ \cite{Amhis:2012bh}:
\begin{eqnarray}
{\cal B }(B_s \to \pi^+ \pi^-)&=&(0.73 \pm 0.14) \times 10^{-6}.
\end{eqnarray}
The branching fraction of another pure annihilation decay  mode $B_d \to K^+ K^-$ has been also measured as~\cite{Amhis:2012bh}:
\begin{eqnarray}
{\cal B }(B_d \to K^+ K^-)&=(0.12 \pm 0.06) \times 10^{-6}
\end{eqnarray}

Theoretically, within  QCD factorization (QCDF) approach \cite{Beneke:1999br},  only an order of magnitude estimate can be given for these two decays  through introducing new phenomenological parameters ($\rho_A$ and $\phi_A$) or an effective gluon propagator \cite{yang} due to the existence of the endpoint singularity. The predicted branching fractions are at the order of $10^{-8}$ \cite{Beneke:1999br, Chang:2014yma}. Moreover, the effects of SU(3) asymmetry breaking  have also been discussed in \cite{Wang:2013fya}. On the contradiction, the perturbative QCD (PQCD) approach \cite{Keum:2000ph}   retains   the transverse momenta of all inner quarks, and thus the endpoint singularity disappear. This  makes the perturbative  calculations of pure annihilation decay modes reliable. On the basis of PQCD, the decays $B_s  \to \pi^+ \pi^-$ and $B_d \to K^+ K^-$   have   been explored in Refs.~\cite{Li:2004ep, Ali2007} and \cite{Chen:2000ih}, respectively. In Ref.~\cite{Xiao:2011tx}, the authors have revisited these two decays with new parameters (especially for the distribution amplitudes of light mesons), and the obtained results are in agreement with the experimental data well. Despite the agreement, by comparing  the predictions of \cite{Xiao:2011tx} with the experimental result, one can find that the LHCb measurement   has a central value larger than the theoretical result, which may indicate some room left for survival of a light $Z^\prime$ boson. In the following we will use the PQCD approach and  investigate  the impact of the family non-universal leptophobic $Z^\prime$ model on  the $CP$ asymmetries of these two decays. Our results can  be stringently tested at  the LHCb experiment, Belle-II, and future high energy $e^+e^-$ collider.

This paper is organized as follows. In Sec.\ref{sec:2}, after  a brief introduction to the PQCD approach, we will present the numerical results of $B_s \to \pi^+ \pi^-$ and $B_d \to K^+ K^-$ in SM. In Sec.\ref{sec:3}, we will discuss the effects of the $Z^\prime$ on the branching fractions and $CP$ asymmetries of these two decay modes. At last, the conclusion will be drawn in the Sec.\ref{sec:4}.

\section{SM Calculation} \label{sec:2}
In this section, we will start with the effective weak Hamiltonian for the  $b\to D$ ($D=d,s$) transitions, which are given by \cite{Buchalla:1995vs}
 \begin{eqnarray}
 {\cal H}_{eff} = \frac{G_{F}}{\sqrt{2}} \bigg\{ \sum\limits_{q=u,c} V_{qb} V_{qD}^{*} \left(C_{1}  O^{q}_{1} + C_{2} O^{q}_{2}\right )- V_{tb} V_{tD}^{*}  {\sum\limits_{i=3}^{10}} C_{i}  O_{i}  \bigg\}+ \mbox{H.c.} ,
 \label{eq:hamiltonian}
\end{eqnarray}
where $V_{qb(D)}$ are the Cabibbo-Kobayashi-Maskawa (CKM) matrix elements. The explicit expressions of the local four-quark operators $O_{i}$ ($i=1,...,10$) and the corresponding wilson coefficients $C_i$ at different scales have been given in Ref.~\cite{Buchalla:1995vs}. Note that $O^q_{1,2}$ are tree operators and others $O_{3-10}$ are penguin ones.

The PQCD approach is based on the $k_T$ factorization,  and has been applied to calculate the non-leptonic $B$ meson decays for many years \cite{Keum:2000ph,Li:2004ep}. In this approach, the decay amplitude is conceptually written as{\small
\begin{eqnarray} \label{eq:pqcd}
{\cal A}\sim
 \int dx_1 dx_2dx_3 \int b_1db_1b_2db_2b_3db_3 {\rm Tr}\Big[C(t) \Phi_B(x_1,b_1) \Phi_2(x_2,b_2) \Phi_3(x_3,b_3) H(x_i,b_i,t)S_t(x_i)e^{-S(t)}\Big],
\end{eqnarray}}
where $x_i$ are the momentum fractions taken by light quarks in each mesons, and $b_i$ are the conjugate variables of the transverse momenta of light quarks. ``Tr" means the trace over both Dirac and color indices. In light of the factorization hypothesis, the wilson coefficient $C(t)$ encapsulates the    dynamics from $m_W$ down to the scale $t$, where $t\sim O(M_B/2)$  is the typical scale of the concerned annihilation type decays. The hard part $H$, involving the four-quark operators and the hard gluon, describes the hard dynamics characterized by the scale $t$, and it can be calculated perturbatively. The wave function $\Phi_M$, standing for hadronization of the quark and anti-quark into the meson $M$, is independent of the specific processes and thus universal. The  factor $S_t(x_i)$ arises  from the resummation of the large double logarithms ($\ln^2 x_i$) on the longitudinal direction, while the Sudakov form factor $e^{-S(t)}$ is from the resummation of the double logarithm $\ln^2 k_T$. Fortunately, the endpoint could be smeared effectively with the help of these two functions, which makes our calculation reliable.

In particular, the wave functions $\Phi_{M,\alpha\beta}$ ($\alpha,\beta$ being Dirac indices) are decomposed in terms of the spin structure, $1_{\alpha\beta}$, $\gamma^\mu_{\alpha\beta}$, $(\gamma_5\sigma^{\mu\nu})_{\alpha\beta}$, $(\gamma^\mu\gamma_5)_{\alpha\beta}$ and  $(\gamma_5) _{\alpha\beta}$. For the heavy pseudo-scalar meson $B_q$ ($q=d,s$) meson, the wave function  $\Phi_{B,\alpha\beta}$ is given by
\begin{eqnarray}
\Phi_{B,\alpha\beta}(x,b)=\frac{i}{\sqrt {2 N_c}} \left\{ (p\sl_B \gamma_5)_{\alpha\beta}+m_B\gamma_{5\alpha\beta})\right\}\phi_B(x,b),
\end{eqnarray}
where $N_c = 3$ is the color degree of freedom, and $p_B$ is the momentum of $B$ meson. The scalar distribution amplitude $\phi_B$ is normalized by its own decay constant $f_B$
\begin{eqnarray}
\int_0^1\phi_B(x,b=0)dx=\frac{f_B}{2\sqrt{2 N_c}} .
\end{eqnarray}
In this work, we employ the function
\begin{eqnarray}
\phi_{B}(x,b)=N_{B}x^{2}(1-x)^{2}\exp \left[ -\frac{1}{2} \left( \frac{xm_{B}}{\omega _{B}}\right) ^{2} -\frac{\omega_{B}^{2}b^{2}}{2}\right]\label{bw} \;,
\end{eqnarray}
where the shape parameter $\omega_{B_d}=0.4$ GeV ($\omega_{B_s}=0.45$ GeV) has been adopted in all  previous analysis of exclusive $B_{d(s)}$ meson decays \cite{Keum:2000ph,Li:2004ep,Ali2007}.

\begin{figure}[tb]
\begin{center}
\psfig{file=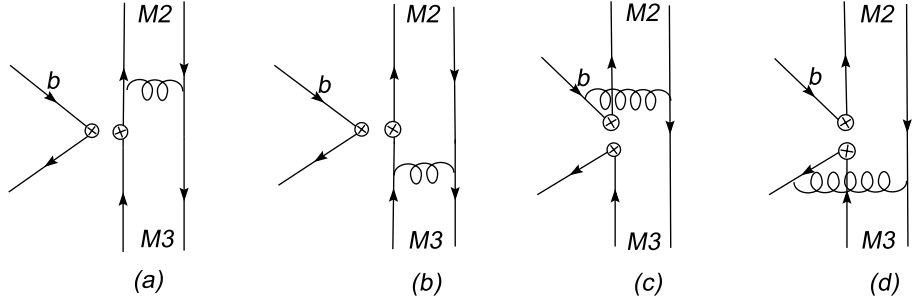,width=12.0cm,angle=0}
\end{center}
\caption{The Feynman diagrams for annihilation
contribution, with possible four-quark operator
insertions}\label{fig:1}
\end{figure}

In contrast to the heavy meson, the wave functions of light meson $\phi_M$ are much complicated due to the non-negligible chiral mass.  Taking the $K^+$ meson as an example for illustration, we define its wave function as
\begin{eqnarray}
\Phi_{K^+,\alpha\beta}(x,b)=\frac{i}{\sqrt {2 N_c}} \Big[ ( \gamma_5p\sl_K)\phi_K^A(x,b)+m_{0K}\gamma_{5}\phi_K^P(x,b)+ m_{0K}\gamma_{5}(v\sl n\sl-1)\phi_K^T(x,b)\Big]_{\alpha\beta},
\end{eqnarray}
where $p_K$ is its momentum, and $m_{0K}=m_K^2/(m_u+m_s)$ is the aforementioned chiral mass. $\vec v$ and $\vec n$ are unit vectors,  and $\vec v $ ($\vec n$) is (anti-)parallel to ${\vec p}_K$.  As nonperturbative parameters, the light cone distribution amplitudes (LCDAs) $\phi^{A,P,T}_M$,  should be fixed by experimental data in principle. Though there is no direct experimental measurement for the moments yet up to now, the non-leptonic charmless $B_q$ decays already give much hints on them \cite{Keum:2000ph,Li:2004ep}. Since the PQCD approach had already given very good results for these decays, especially for the direct $CP$ asymmetries in $B^0\to \pi^+\pi^-$ and $B^0 \to K^+\pi^-$ decays, we shall adopt the well constrained LCDAs of the mesons in these papers~\cite{PseudoscalarWV} (see \cite{Pseudoscalar-Ball} for a summary and update of the LCDAs):
\begin{eqnarray}
 \phi_{\pi}^A(x) &=& \frac{3f_{\pi}}{\sqrt{6}} x(1-x)[1 +0.44C_2^{3/2}(t)], \nonumber \\
 \phi_{\pi}^P(x) &=& \frac{f_{\pi}}{2\sqrt{6}}[1 +0.43C_2^{1/2}(t) ],\nonumber \\
 \phi_{\pi}^T(x) &=& -\frac{f_{\pi}}{2\sqrt{6}}[C_1^{1/2} (t)+0.55 C_3^{1/2} (t)]
 ,\nonumber\\
 \phi_{K}^A(x) &=& \frac{3f_{K}}{\sqrt{6}}x(1-x)[1+0.17C_1^{3/2}(t)+0.2C_2^{3/2}(t)], \nonumber\\
 \phi_{K}^P(x) &=& \frac{f_{K}}{2\sqrt{6}} [1+0.24C_2^{1/2}(t)],\nonumber \\
 \phi_{K}^T(x)
 &=&-\frac{f_{K}}{2\sqrt{6}}[C_1^{1/2} (t)+0.35 C_3^{1/2} (t)],
\end{eqnarray}
with Gegenbauer polynomials defined as:
\begin{eqnarray}
 C^{1/2}_{1}(t)=t,  C^{3/2}_{1}(t)=3t,
 C_2^{1/2}(t)=\frac{1}{2} (3t^2-1),  C_2^{3/2}(t)=\frac{3}{2}
(5t^2-1),  C_3^{1/2}(t)=\frac{1}{2} t (5t^2 -3),
\end{eqnarray}
and $t=2x-1$. It should be stressed that we have dropped the terms proportional to  $C_4^{1/2,3/2}$, and only kept the first two terms, following the arguments of \cite{Ali2007}.

Now we turn to calculate the hard part $H$. According to the effective Hamiltonian, eq.(\ref{eq:hamiltonian}), we can draw four kinds of Feynman diagrams contributing to the $B_d \to K^+ K^-$ and $B_s \to \pi^+ \pi^-$ decays at the leading order, as is shown in Fig.\ref{fig:1}. The four diagrams are classed into two types: ($a$) and ($b$) are factorizable diagrams , and ($c$) and ($d$) are non-factorizable ones. Due to the current conservation, the contributions from the factorizable diagrams $(a)$ and $(b)$ will be canceled exactly by each other, so that contributions from diagrams (a) and (b) are null. As concerned as diagrams ($c$) and ($d$), by inserting the possible operators, we can obtain the amplitudes for the non-factorizable annihilation diagram $M_{ann}^{LL}$ and $M_{ann}^{SP}$, where $LL$ stands for the contribution from $(V-A)(V-A)$ operators, and $SP$ for the contribution from $(S-P)(S+P)$ operators which result from the Fierz transformation of the $(V-A)(V+A)$ operators. The expressions of $M_{ann}^{LL}$ and $M_{ann}^{SP}$ and the inner functions can be found in \cite{Ali2007}. Finally, we obtain total decay amplitudes for concerned decays as
\begin{eqnarray}
{\cal A}(\overline B_{s}\to\pi^{+}\pi^{-}) &=& \frac{G_F}{\sqrt{2}} \Big\{V_{ub}V_{us}^{*} M_{ann}^{LL}\left[C_2\right]- V_{tb}V_{ts}^{*}\Big(M_{ann}^{LL}\left[ C_{4}+C_{10}\right]+M_{ann}^{SP}\left[C_6+C_{8}\right]  \nonumber \label{bspipi}\\
&& +M_{ann}^{LL}[C_{4}-\frac{1}{2}C_{10}]_{\pi^-\leftrightarrow \pi^+}+ M_{ann}^{SP}[ C_6-\frac{1}{2}C_6]_{\pi^-\leftrightarrow \pi^+}\Big)\Big\};\\
{\cal A}(\overline B_{d}\to K^{+}K^{-}) &=& \frac{G_F}{\sqrt{2}} \Big\{V_{ub}V_{ud}^{*} M_{ann}^{LL}\left[C_2\right] -V_{tb}V_{td}^{*}\Big(M_{ann}^{LL}\left[ C_{4}+C_{10}\right]+M_{ann}^{SP}\left[C_6+C_{8}\right]   \nonumber\\
&&  +M_{ann}^{LL}[C_{4}-\frac{1}{2}C_{10}]_{K^-\leftrightarrow K^+}+M_{ann}^{SP}[C_6-\frac{1}{2}C_8]_{K^-\leftrightarrow K^+}\Big)
\Big\}.\label{bdKK}
\end{eqnarray}
In eq.(\ref{bspipi}), when $\pi^+$ and $\pi^-$ exchanging, $M_{ann}^{LL}$ obtain the results because of SU(2) symmetry. Furthermore, if we ignore the small $x_1$ (the momentum fraction of $s$ quark in the $\overline B_s$ meson) in the denominators, $M_{ann}^{LL}$ is same as $M_{ann}^{SP}$,too \footnote{In Ref.\cite{Xiao:2011tx}, there are typos in eqs.(27) and (28).}. However, for $\overline B_d \to K^+K^-$, $M_{ann}^{LL(SP)}$ do not share same formulaes with $M_{ann}^{LL(SP)}|_{K^-\leftrightarrow K^+}$ due to the difference between the mass of up (down) quark and that of the strange quark, and such difference might affect the direct $CP$ asymmetry. In fact, in our calculations there are many uncertainties,  the most important one of which is from the distribution amplitude of initial heavy meson, because it cannot calculated directly from QCD till now yet. In the following work, we shall vary the shape parameter $\omega_{B_d}=0.40\pm 0.05$ and $\omega_{B_s}=0.50\pm 0.05$. Furthermore, the contributions from next leading order (NLO) have not been done. In the current work, to estimate the uncertainties of NLO , we simply vary $t$ from $0.8t$ to $1.2t$, where $t$ is the largest scale in each diagram and the expressions of them have been given in \cite{Ali2007}. Combining all above uncertainties, we obtain the $CP$-averaged branching fractions of two decay modes
\begin{eqnarray}
{\cal B} (B_s\rightarrow \pi^+\pi^-)&=&(5.5^{+1.1}_{-0.9})\times 10^{-7}.\\
{\cal B} (B_d\rightarrow K^+K^-)&=&(1.9^{+0.3}_{-0.3})\times 10^{-7}.
\end{eqnarray}
Since the uncertainties from the $\pi, K$ meson distribution amplitudes are very small, we will not discuss them here.

In discussing the $B$ meson decays, we usually define direct $CP$ asymmetry as
\begin{eqnarray}
A_{CP}^{dir}\equiv \frac{|A(\overline B_q\to f)|^2-|A(B_q\to \overline f)|^2}{|A(\overline B_q\to
f)|^2+|A(B_q\to \overline f)|^2}.\label{ACPDef}
\end{eqnarray}
Moveover, because the final states $\pi^+\pi^-, K^+K^-$ have definite $CP$-parity, one can measure the time-dependent decay width of the the $B_q \to f$ decay \cite{BsmixingDun}:
\begin{eqnarray}
\Gamma(B(t)\to f)\propto \cosh \Big(\frac{\Delta \Gamma t}{2}\Big)
+H_f \sinh \Big(\frac{\Delta \Gamma t}{2}\Big)
-{\cal A}_{\mathrm{CP}}^{dir}\cos( \Delta m t)- S_f \sin (\Delta mt),
\end{eqnarray}
where $\Delta m=m_H-m_L>0$ is the mass difference, and  $\Delta \Gamma=\Gamma_H-\Gamma_L$ is the difference of decay widths for the heavier and lighter $B_q^0$ mass eigenstates. Correspondingly, the time dependent decay width $\Gamma(\overline B^0_q(t)\to f)$ is obtained from the above expression by flipping the signs of the $\cos(\Delta m t)$ and $\sin(\Delta m t)$ terms. The  $S_f$ and $H_f$ that can be extracted from the-time dependent decay width are defined as
\begin{eqnarray}
S_f \equiv \frac{2Im[\lambda]}{1+|\lambda|^2}, ~~~ H_f \equiv\frac{2Re[\lambda]}{1+|\lambda|^2},
\end{eqnarray}
with
\begin{equation}
\lambda=\eta_f e^{2i\epsilon}\frac{A(\overline B_q\to f)}{A(B_q\to\bar f)},
\end{equation}
where  $\eta_f$ is $+1(-1)$ for a CP-even (CP-odd) final state $f$ and $\epsilon=\mbox{arg}[-V_{cb}V_{tq}V^*_{cq}V^*_{tb}]$. In SM, the predicted results are listed as
\begin{eqnarray}
\left\{
  \begin{array}{ll}
    {\cal A}_{CP}^{dir} (B_s\rightarrow \pi^+\pi^-)=(-1.5\pm0.2)\%, &   \\
    {\cal S}_{f} (B_s\rightarrow \pi^+\pi^-)=0.11\pm0.01, &   \\
    {\cal H}_{f} (B_s\rightarrow \pi^+\pi^-)=0.99; &
  \end{array}
\right.\\
\left\{
  \begin{array}{ll}
    {\cal A}_{CP}^{dir} (B_d\rightarrow K^+K^-)=(37^{+5}_{-7})\%, &   \\
    {\cal S}_{f} (B_d\rightarrow K^+K^-)=-0.81\pm 0.05, &   \\
    {\cal H}_{f} (B_d\rightarrow K^+K^-)=-0.45\pm 0.05, &
  \end{array}
\right.
\end{eqnarray}

For $B_s\rightarrow \pi^+\pi^-$, both branching fraction and $CP$ asymmetry parameters agree with previous studies \cite{Li:2004ep,Ali2007,Xiao:2011tx}, and small differences are from the uncertainties of  the CKM matrix elements. For $B_d\rightarrow K^+K^-$, our branching fraction consist with prediction of \cite{Xiao:2011tx}, but the direct $CP$ asymmetry is much smaller than theirs because they may omitted the effect of the SU(3) asymmetry in the LCDAs of $K$ meson. Compared with the experimental data, although our branching fractions are consistent with data after considering the uncertainties of both theoretical and experimental sides, the center value of $B_d \to K^+K^-$ ($B_s \to \pi^+\pi^-$) is a bit larger (smaller) than the data, which means there is a little room for us to search for possible effect of NP. Unfortunately, the  $CP$ asymmetries of these two decays have not been measured in the current experiments.

\section{The Contribution of The $Z^\prime$ Boson}\label{sec:3}
Now we shall study the possible contributions of the extra gauge boson $Z^\prime$ in these two decay modes. Ignoring the interference between $Z$ and $Z^\prime$ bosons, we write the Lagrangian with $Z^\prime$ on the gauge interaction basis as
\begin{eqnarray}\label{zprimed}
{\cal L}^{Z^\prime}=-g_2 Z^{\prime {\mu}}\sum_{i,j} {\overline \psi_i^I} \gamma_{\mu}
  \left[ (\epsilon_{\psi_L})_{ij} P_L + (\epsilon_{\psi_R})_{ij} P_R \right] \psi^I_j ~,
\end{eqnarray}
where the field $\psi_i$ stands for the $i$th family fermion, $g_2$ for the coupling constant, $\epsilon_{\psi_L}$ ($\epsilon_{\psi_R}$) for the left-handed (right-handed) chiral coupling, and $P_{L,R}=(1\mp\gamma_5)/2$. When rotating to the physical basis, the mass eigenstates will be obtained by $\psi_{L,R} = V_{\psi_{L,R}} \psi_{L,R}^I$, and the usual CKM matrix is given by $V_{\rm CKM} = V_{u_L} V_{d_L}^{\dagger}$. Similarly, we can get the coupling matrices in the physical basis of up (down)-type quarks,
\begin{eqnarray}
B^X_u \equiv V_{u_X} \epsilon_{u_X} V_{u_X}^{\dagger} ~, ~~ B^X_d
\equiv V_{d_X} \epsilon_{d_X} V_{d_X}^{\dagger} ~~ (X = L,R).
\end{eqnarray}
It is apparent that if $\epsilon_{u(d)_{L(R)}}$ are not proportional to the identity matrix, the nonzero off-diagonal elements in the $B^{L,R}_{u,d}$ will appear, which induce the FCNC interactions at the tree level. In the current work,we assume that the up-type coupling matrix $\epsilon_{u_{L(R)}}$ are proportional to the unit matrix, and the right-handed couplings of are flavor-diagonal for simplicity.  Thereby, the effective Hamiltonian mediated by the $Z^\prime$, for example $ b \to s {\bar q}q~(q=u,d)$ transition, is given by
\begin{eqnarray}
{\cal H}_{\rm eff}^{Z'}=\frac{2 G_F}{\sqrt{2}} \left(\frac{g_2
m_Z}{g_1 m_{Z'}}\right)^2 B^{L}_{bs}({\bar s }b)_{V-A} \sum_q \left( B^L_{qq} ({\bar q}q)_{V-A}   + B^R_{qq} ({\bar q}q)_{V+A} \right) + \mbox{h.c.} ~,
\label{eqn:Heff1}
\end{eqnarray}
where $g_1=e/(\sin{\theta_W}\cos{\theta_W})$ and $m_{Z^{\prime}}$ is the mass of $Z^\prime$ boson. The diagonal elements of the effective coupling matrices $B_{qq}^{L,R}$ are required to be real because of the hermiticity of the effective Hamiltonian. However, for the off-diagonal one of $B_{bs}^{L}$, it might be a complex number and a new weak phase $\phi_{bs}$ is introduced, which might play important roles in explaining the large $CP$ asymmetries in $B \to K\pi$ \cite{Chang:2014yma}. Since the above operators of the forms $({\bar s}b)_{V-A} ({\bar q}q)_{V\mp A}$ already exist in SM, we can represent the $Z^\prime$ effect by modifying the wilson coefficients of the corresponding operators. As a result, we reorganize the eq.(\ref{eqn:Heff1}) as
\begin{eqnarray}
{\cal H}_{\rm eff}^{Z^\prime}=-\frac{G_F}{\sqrt{2}} V_{tb}V_{ts}^*\sum_q \left( \Delta C_3^s O_3^{(q)} +\Delta C_5^s O_5^{(q)} + \Delta C_7^s O_7^{(q)} + \Delta C_9^s O_9^{(q)} \right) +\mbox{h.c.}. \label{eqn:Heff2}
\end{eqnarray}
Correspondingly, the contributions of the extra $Z^\prime$ boson to the SM wilson coefficients at
the $m_W$ scale is given
\begin{eqnarray}
&& \Delta C^s_{3(5)}=- \frac{2}{3 V_{tb}V_{ts}^* }\left(\frac{g_2 m_Z}{g_1m_{Z^\prime}}\right)^2 B^{L}_{bs} \left(B^{L(R)}_{uu} + 2 B^{L(R)}_{dd}\right),\\
&& \Delta C^s_{9(7)} = -\frac{4}{3 V_{tb} V_{ts}^*} \left(\frac{g_2m_Z}{g_1 m_{Z^\prime}}\right)^2 B^{L}_{bs} \left(B^{L(R)}_{uu} - B^{L(R)}_{dd} \right).
\end{eqnarray}
One can see that the $Z^\prime$ contributes to the electro-weak penguins $\Delta C_{9(7)}$ as well as the QCD penguins $\Delta C_{3(5)}$. In order to show that the new physics is primarily manifest in the electro-weak penguins, we simply assume $B^{L(R)}_{uu} \simeq -2 B^{L(R)}_{dd}$, and this relation has been used widely \cite{fcncconstraints,chiang1,BZprime,changkllzp}. Therefore, the $Z'$ contributions to the wilson coefficients are
\begin{eqnarray}
\Delta C_{3(5)}^s= 0,\,\,\,\,\,\,\,\,\,
\Delta C_{9(7)}^s = 4 \frac{|V_{tb}V_{ts}^*|}{V_{tb}V_{ts}^* }
\zeta_s^{LL(R)} e^{i\phi_{bs}} ~,
\end{eqnarray}
where
\begin{eqnarray}\label{eqn:xi}
\zeta^{LX}_s=\left(\frac{g_2 m_Z}{g_1 m_{Z'}}\right)^2
\left|\frac{B^{L}_{bs} B^X_{dd}}{V_{tb}V_{ts}^* }\right| ~~
(X=L,R), \,\,\,\,\,\,\,\,\,
\phi_{bs} ={\rm Arg}[B^L_{bs}] ~.
\end{eqnarray}
Note that the other SM wilson coefficients at scale lower than $m_b$ will also receive contributions from the $Z^{\prime}$ boson through renormalization group (RG) evolution. Since in this model there is no new particle below $m_W$, the RG evolution of the modified Wilson coefficients is exactly  same as the one  in SM~\cite{Buchalla:1995vs}.

Similarly, we also obtain the hamiltonian $b\to d \bar q q$, and the corresponding wilson coefficients and inner functions are given as
\begin{eqnarray}
&& \Delta C_{3(5)}^d= 0,\,\,\,\,\,\,\,\,\,
\Delta C_{9(7)}^d = 4 \frac{|V_{tb}V_{td}^*|}{V_{tb}V_{td}^* }
\zeta_d^{LL(R)} e^{i\phi_{bd}} ~, \\
&&\zeta^{LX}_d=\left(\frac{g_2 m_Z}{g_1 m_{Z'}}\right)^2
\left|\frac{B^{L}_{bd} B^X_{ss}}{V_{tb}V_{td}^* }\right| ~~
(X=L,R), \,\,\,\,\,\,\,\,\,
\phi_{bd} ={\rm Arg}[B^L_{bd}] ~.
\end{eqnarray}

Now, we are in a position to discuss the possible parameter spaces of $\zeta^{LL,LR}_{s,d}$ and $\phi_{bs,bd}$. In particular, we assume $g_2/g_1 \sim 1$ because we expect that both the hypercharge $U(1)_Y$ gauge group and the extra $U(1)^\prime$ have the same origin from some grand unified models. Furthermore, we also hope $ m_Z/m_{Z^\prime}$ is at the order of ${\cal O}(10^{-1})$, so that the neutral $Z^\prime$ boson could be detected at LHC experiment directly. Note that the mass of a leptophobic $m_{Z^\prime}$ boson has not been constrained till now, as aforementioned in Sec.\ref{sec:1}. In addition, we need to determine the other parameters $|B^{L}_{bs}|$, $|B^{L}_{bd}|$, $|B^X_{qq}|$ and new weak phases $\phi_{bd,bs}$ with the accurate data from $B$ factories and LHCb experiment. For example, $B^L_{bs,bd}$ and $\phi_{bs,bd}$ could be extracted from $B_q^0$-$\overline B_q^0$ ($q=d, s$) mixing. In order to explain the mass differences between $B_{q}^0$ and $\overline B_{q}^0$ with new $Z^\prime$ boson, $|B^{L}_{bs(d)}|\sim|V_{tb}V_{ts(d)}^{*}|$ is required. Then, with experimental data of $B_{d,s}$ nonleptonic charmless decays, $B^{L,R}_{qq}\sim 1$ could be extracted. For the new introduced phases $\phi_{bs}$ and $\phi_{bd}$, they have not been constrained totally, although many efforts have been done \cite{changkllzp}, we  therefore set them as free parameters. How to constraint of these parameters globally is beyond the scope of current work and can be found in many references \cite{chiang1, BZprime,changkllzp}. So as to probe the new physics effect for maximum range, we assume  $\zeta \sim \zeta^{LL}_{d,s}\sim\zeta^{LR}_{d,s}\in [0.001,0.02]$, ie, the range of $m_{Z^\prime}$ is about $[636, 2800]~\mathrm{GeV}$, and $\phi_{bd,bs} \in [-180^\circ, 180^\circ]$.

\begin{figure}[tb]
\begin{center}
\psfig{file=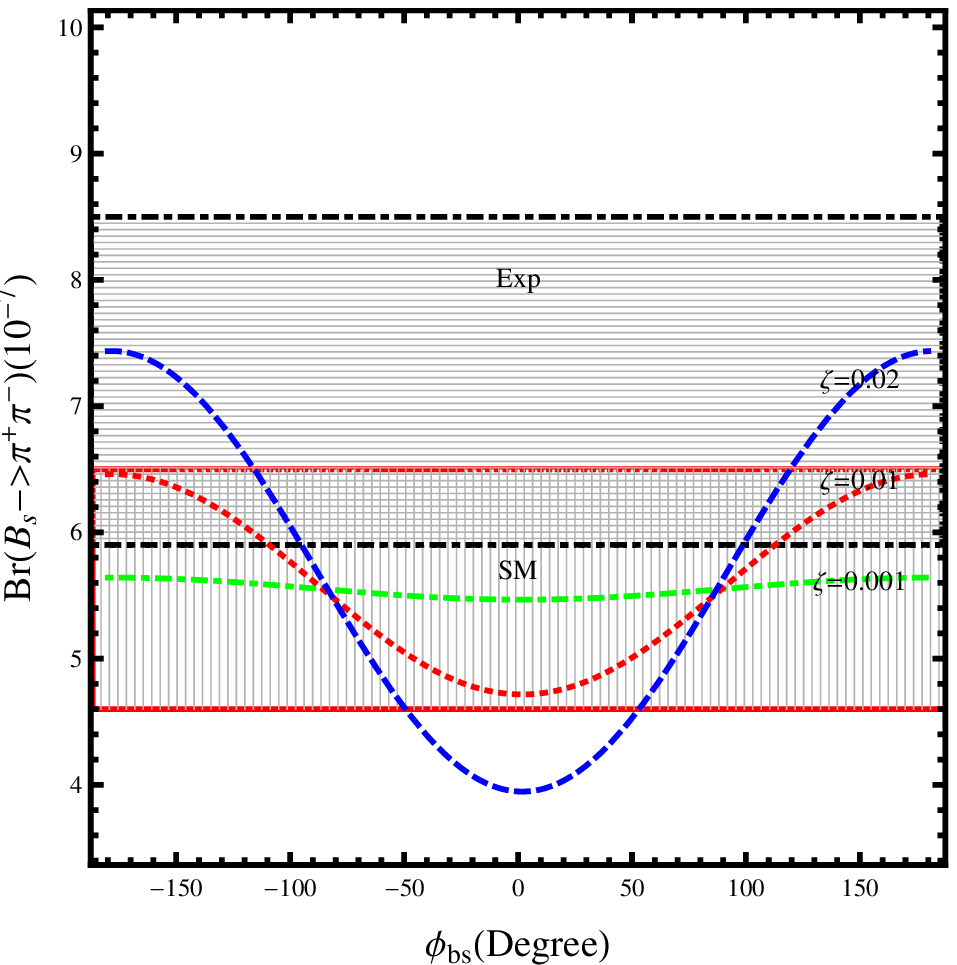,width=6.0cm,angle=0}
\hspace{2cm}
\psfig{file=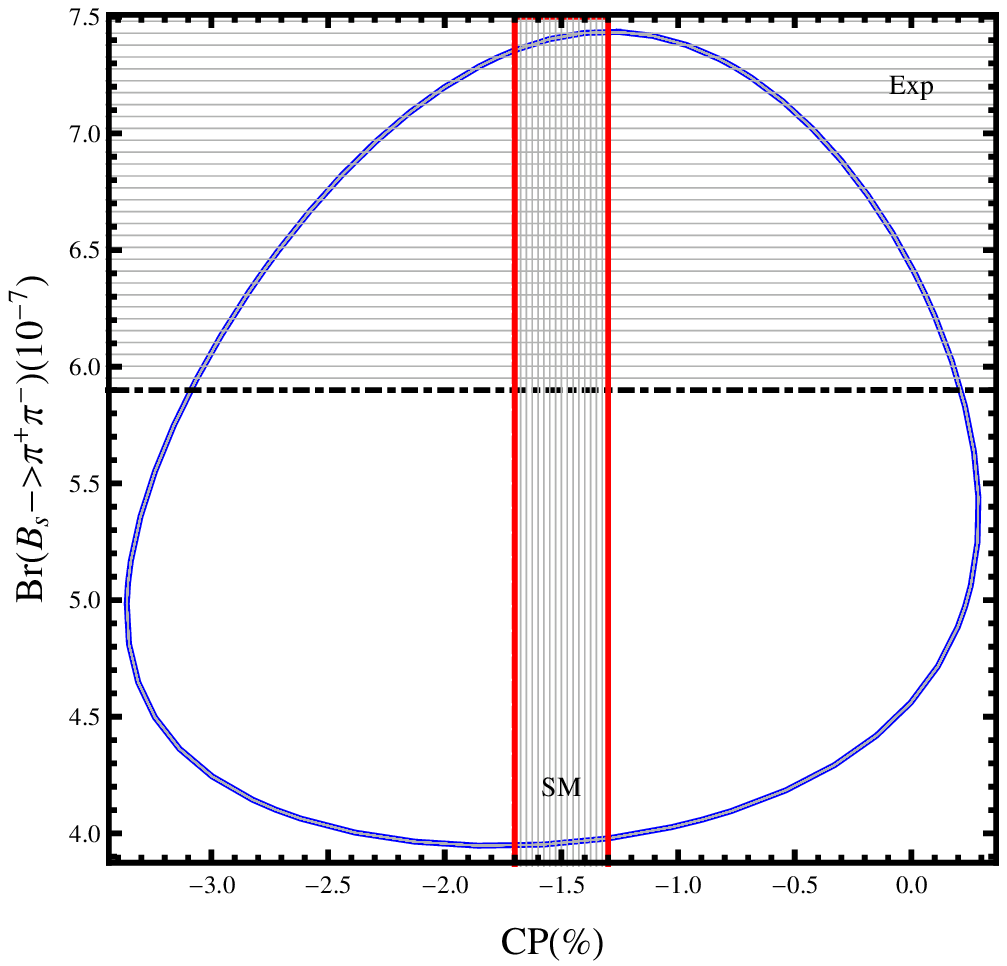,width=6.35cm,angle=0}
\end{center}
\caption{The contribution of $Z^\prime $ to the decay mode $B_s \to \pi^+\pi^-$. The left panel represents  the branching fraction as functions of $\phi_{bs}$, the dotdashed (green), dotted (red), and dashed (blue) lines represent results from the $\zeta=0.001,~0.01,~0.02$, respectively. The region edged by dotdashed lines (black) is the experimental data, while edged by the solid lines (red) is prediction of SM. The right panel stands for the relation between the branching fraction and the direct $CP$ asymmetry,the region edged by dotdashed lines (black) is the experimental data, while edged by the solid lines (red) is prediction of SM. }\label{Fig:bspipi}
\end{figure}

In Fig.\ref{Fig:bspipi}, we explore the possible effects of $Z^\prime$ boson on  the decay mode $B_s \to \pi^+\pi^-$. In the left panel, we present the variation of the CP-averaged branching fraction as a function of the new weak phase $\phi_{bs}$ with different $\zeta = 0.001(\mathrm{dotdashed}), 0.01 (\mathrm{dotted}), 0.02(\mathrm{dashed})$. The experimental region (filled by horizontal lines) and the SM predictions (filled by vertical lines) are also shown for the comparison. From this figure, one can see that  the SM is consistent with the data within $1\sigma$. Including  the $Z^\prime$ contribution, it is apparent that the parameter space $|\phi_{bs}|< 80^\circ$ will be excluded. For $|\phi_{bs}|> 80 ^\circ$, if $\zeta< 0.01$, the contribution of $Z^\prime$ boson will be buried by the uncertainties of SM. One also sees that when $\zeta= 0.02$ the branching ratio will be enhanced to $7.6 \times 10^{-7}$, which is larger than the SM prediction. Note that the averaged experimental have large errors, and the small band will help us to determine the magnitude of $\zeta$. It is emphasized that the LHCb had obtained a bit larger result, which indicates the existence of a light $Z^\prime$. In the right panel, we plot the relation between the branching ratio and the direct $CP$ asymmetry ${\cal A}_{CP}^{dir}$ with an extra $Z^\prime$ boson. The region edged by blue curve is the possible region with parameter $\zeta<0.02$ and $\phi_{bs} \in[-180^\circ, 180^\circ]$. With the experimental data, the lower half of the region can be excluded. In the permitted region, the range of direct $CP$ asymmetry is $[-3.2\%,0.1\%]$, which is much larger than the estimation of SM (the grey region). The future measurement of these values in LHCb (LHC-II) experiment and the high energy $e^+e^-$ collider will help us to probe the effects of $Z^\prime$. Note that when discussing the effects of $Z^\prime$ boson, we will not include the uncertainties induced by wave functions and scale $t$, because the major objective of this work is searching for the possibility of new physics signal, rather than producing acute numerical results.

\begin{figure}[tb]
\begin{center}
\psfig{file=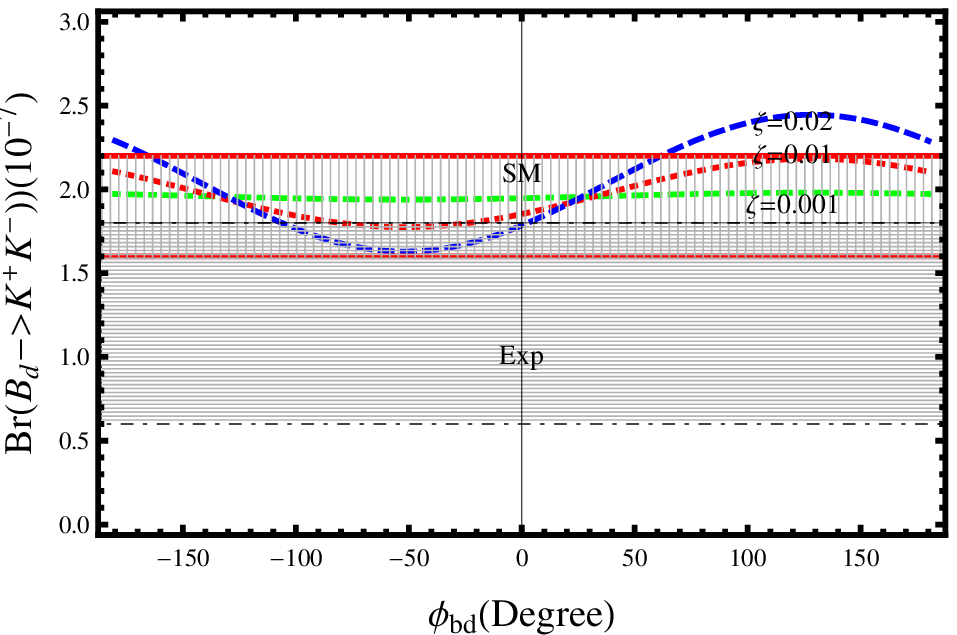,width=5.5cm,height=5.2cm}
\hspace{2cm}
\psfig{file=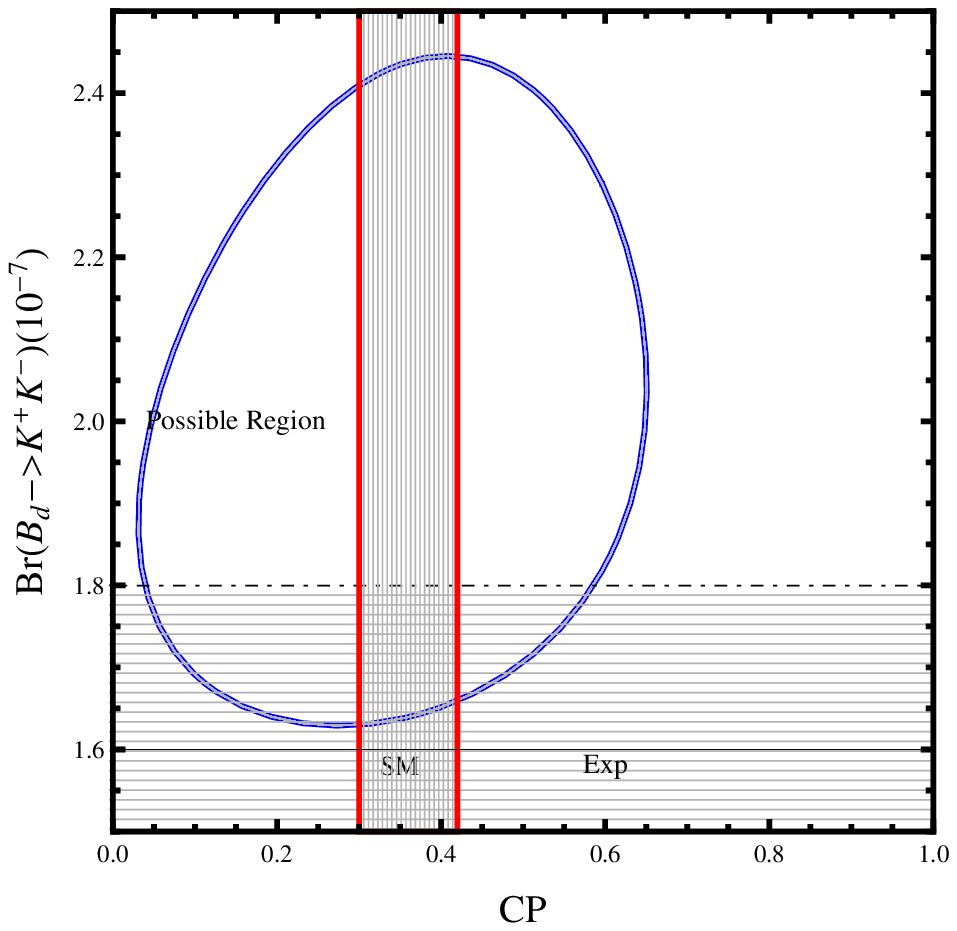,width=6cm,height=5cm}
\end{center}
\caption{The contribution of $Z^\prime $ to the decay mode $B_d \to K^+K^-$. The legends are the same as in Fig. \ref{Fig:bspipi}}\label{Fig:bdKK}
\end{figure}

Similarly, the effects of the extra $Z^\prime$ boson in $B_d \to K^+K^-$ are also presented in Fig. \ref{Fig:bdKK}. In the left panel, it is clear that the position of SM prediction is on the top of the experimental data, although some parts of them overlap with each other. Furthermore, the heavy $Z^\prime$ contributions ($\zeta<0.1$) are not apparent due to the uncertainties of SM.  Moreover, for the $\phi_{bd}$, the ranges $[0^\circ, 180^\circ]$ and $[-180^\circ,-110^\circ]$ will be excluded.  We also plot the region (edged by curve) related to the direct $CP$ asymmetry and the branching fraction, as shown in the right panel. Note that the SM prediction is $30\% \sim 42\%$, but with a light  $Z^\prime$ the estimated range is to be $5\% \sim 60\%$ after considering the constraint from the experimental data. It is concluded that for $B_d \to K^+K^-$ the future measurement of direct $CP$ asymmetry will help us to search for the possible effect of a light $Z^\prime$, though its contribution to the branching fraction is polluted by the SM uncertainties.

\begin{figure}[tb]
\begin{center}
\psfig{file=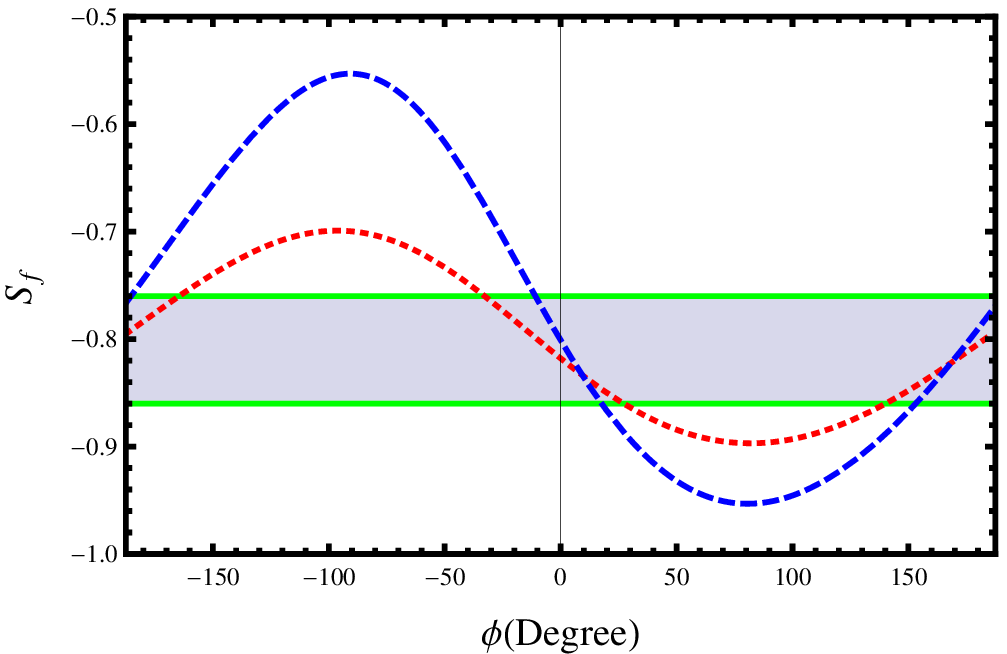,scale=0.7}
\hspace{2cm}
\psfig{file=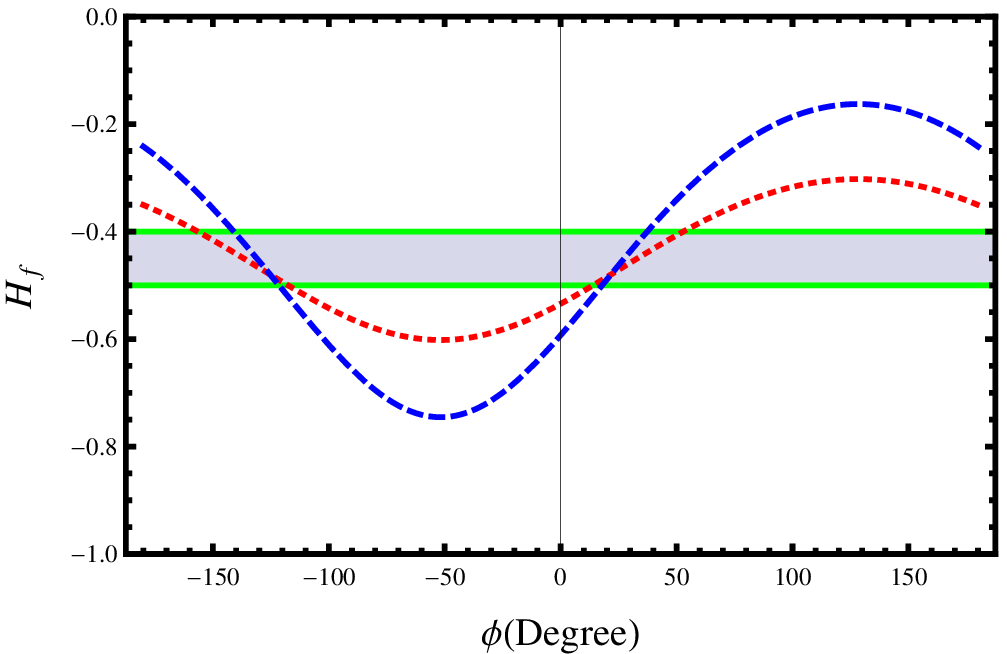,scale=0.7}
\end{center}
\caption{The $CP$ symmetry parameters ${\cal S}_f$ (left panel) and ${\cal H}_f$ (right panel) as a function of the weak phase $\phi_{bd}$, the dotted (red) and dashed (blue) lines represent results from the $\zeta=0.01,~0.02$, and the regions edged by solid line (green) are the predictions of SM.} \label{Fig:bdKK2}
\end{figure}

At last, we shall discuss the $Z^\prime$ effect on the $CP$ symmetry parameters ${\cal S}_f$ and ${\cal H}_f$. For $B_s \to \pi^+\pi^-$, as the weak phase of $V_{tb}V_{ts}^*$ is very small, both ${\cal S}_f$ and ${\cal H}_f$ are not sensitive to the NP. On the contrary, for $B_d \to K^+K^-$,  ${\cal S}_f$ and ${\cal H}_f$ are sensitive to the extra leptophobic $Z^\prime$ boson. In Fig.\ref{Fig:bdKK2}, we plot the relations of ${\cal S}_f$ (right panel) and ${\cal H}_f$ (left panel) with varying $\phi_{bd}$ from $-180^\circ$ to $ 180^\circ$, when $\zeta = 0.01$ and $\zeta = 0.02$. The estimations of SM (edged by the lines) are also presented. From the figures, one can see that in the permitted range of $\phi_{bd}$, with a light $Z^\prime$ boson ($\zeta = 0.02$), ${\cal S}_f$ could reach $-0.55$, which is larger than the prediction of SM. For ${\cal H}_f$, its values could reach to $-0.75$ when $\phi_{bd}=-50^\circ$. The future measurement of them will help us to  further constrain the parameters, which might helpful for direct searching for a light $Z^\prime$ boson.

\section{Summary}\label{sec:4}
In this work, we have studied the pure annihilation decays $B_d \to K^+K^-$ and $B_s \to \pi^+\pi^-$ in the SM and the family non-universal leptophobic $Z^\prime$ model. Although the SM predictions in the PQCD approach are in agreement with experimental data, the branching fraction of $B_s \to \pi^+\pi^-$ ($B_d \to K^+K^-$) is a little bigger (smaller) than the LHCb result, which may indicate the survival  space for  a light $Z^\prime$ boson. Inspired by this fact, we have constrained  the  $U(1)$  phase  as $\phi_{bd} \in [-110^\circ,0^\circ]$, and  $\phi_{bs}$ is $|\phi_{bs}|>80^\circ$. Within the allowed space range, the    direct $CP$ asymmetry for $B_s \to \pi^+\pi^-$ is predicted as  $[-3.2\%,0.1\%]$, while  it is $5\% \sim 60\%$ for $B_d \to K^+K^-$. Furthermore, we  have also calculated the mixing-induced $CP$ asymmetries and found that the parameters ${\cal S}_f$ and ${\cal H}_f$ of $B_d \to K^+K^-$ are very sensitive to the effect of $Z^\prime$. With a light $Z^\prime$, the maximum (minimal) value of ${\cal S}_f $(${\cal H}_f$) can reach $-0.55$ ($-0.75$). The future measurements of these observables may provide us some hints for direct searching for a light leptophobic $Z^\prime$ boson.

\section*{Acknowledgments}
This work is supported by the National Science Foundation (Grants No. 11175151 and No. 11235005), and the Program for New Century Excellent Talents in University (NCET) by Ministry of Education of P. R. China (Grant No. NCET-13-0991).



\end{document}